\documentclass[12pt]{article}
\usepackage[dvips]{graphicx}
\begin{document}
\sf
\begin{center}
   \vskip 2em
    {\LARGE \sf Inverted catenoids, curvature singularities}
\vskip 0.5em
 { \LARGE \sf and tethered membranes}
 \vskip 3em
 {\large \sf  Pavel Castro-Villarreal and Jemal Guven \\[2em]}
\em{ Instituto de Ciencias Nucleares,
 Universidad Nacional Aut\'onoma de M\'exico\\
 Apdo. Postal 70-543, 04510 M\'exico, DF, MEXICO\\[1em]
}
\end{center}
 \vskip 1em

\begin{abstract}
If a catenoid is inverted in any interior point, a deflated compact
geometry is obtained which touches at two points (its poles). The
catenoid is a minimal surface and, as such, is an equilibrium shape
of a symmetric fluid membrane. The conformal symmetry of the
Hamiltonian implies that inverted minimal surfaces are also
equilibrium shapes. However, they exhibit curvature singularities at
their poles. These singularities are associated with external forces
pulling the poles together. Unlike the catenoid which is free of
stress, there will be stress within the inverted shapes. If the
surface area is fixed, reducing the external force induces a
transition from a discocyte to a cup-shaped stomatocyte.

\end{abstract}
 \vskip 3em

\section{\Large \sf Introduction}

Biological membranes are fluid bilayers whose mechanical properties
are described on mesoscopic scales by a bending energy quadratic in
extrinsic curvature \cite{SackLip,Seifert,Mouritsen}. If bilayer
asymmetry or constraints are ignored, this energy is invariant under
the global conformal transformations of three-dimensional space
\cite{Willmore,Polyakov,Seifert1}. Thus the bending energy is not
only independent of the size of the membrane, it is unchanged under
inversion in any point, ${\bf x} \to {\bf x}/ |{\bf x}|^2$. As a
consequence,  every equilibrium configuration has a counterpart
related to it by inversion which is also an equilibrium. Remarkably,
this apparently unnatural symmetry does have physical consequences.

\vskip1pc In practice, what one does is pair inversions in special
conformal transformations which can be treated perturbatively. The
study of inversion in its own right, however, lies beyond the reach
of perturbation theory.  If the point of inversion lies on the
surface, this point gets consigned to infinity; remote points on the
surface, on the other hand get mapped into a neighborhood of the
origin i.e. shapes do not simply get distorted, they may find
themselves distorted beyond recognition; even their topology might
change. Two shapes related by inversion will typically describe very
different physical conditions. Inversion has the potential to
generate curvature singularities where none existed.

\vskip1pc In this paper, we will examine various consequences for
the physics of fluid membranes of this non-perturbative  aspect of
conformal symmetry. To keep things simple we will look only at
axially symmetric shapes. If our objective had been to identify
compact isolated geometries without internal or external
constraints, this symmetry would be a little too confining: the only
equilibrium geometries of this kind are spheres and Clifford torii
with ratio of the wheel to tube radius of $\sqrt{2}$.

\vskip1pc The catenoid is also an equilibrium geometry albeit an
infinite one. Finite sections of catenoid play an important role as
a geometrical component of fluid membranes \cite{SackLip}. Because
mean curvature vanishes, the contribution made by the catenoid to
the bending energy is entirely topological, depending only on the
boundary behavior. The central region of the catenoid provides the
neck geometry permitting membranes to bud without incurring the
large energy penalties one might expect to be associated with the
high curvatures which form in the neck region. While the curvatures
may be large, the mean curvature and thus the added local energy
vanishes.  This is also the geometry involved in the assembly of
topologically complicated membranes. Thus it should not be
surprising that such structures are abundant in biological cells.

\vskip1pc Spheres and torii are not very interesting under
inversion. The catenoid, however, maps not only to a completely
different equilibrium geometry, but to one that is also interesting
physically. In particular, if the catenoid is inverted in its center
of symmetry it transforms into a compact biconcave geometry: a
discocyte. The two distal regions of the catenoid get mapped to the
point of inversion. Thus the north and south poles of this geometry
touch and, because the geometry possesses a tangent plane at these
points, it is appropriate to think of the topology as spherical.

\vskip1pc The problem is: how does one interpret such an
equilibrium?  The only regular equilibrium geometries with spherical
topology are round spheres. Without constraints, one would thus
expect this geometry to equilibrate by inflating into a sphere.
Physically, the only way such a deflated equilibrium could exist is
if there are external forces counteracting the tendency to inflate.
Such forces will act as a source of stress in the membrane
\cite{MDG}. The appropriate way to quantify this connection is to
describe the geometry in terms of a conserved stress tensor
\cite{Stress,Auxiliary}.

\vskip1pc Unlike a catenoid formed by a soap film spanning two rings
which is under tension, a catenoid formed by an ideal fluid membrane
is free of stress. However, the equilibrium described by its
inverted counterpart is very different: it is under stress. As shown
in \cite{guvencastro}, the source of this stress is a pair of equal
and opposite localized external forces at the poles pulling them
together.

\vskip1pc External forces manifest themselves in curvature
singularities. Despite the apparent smoothness of the inverted
catenoid at the poles, there is a singularity lurking at these
points. The two principal curvatures diverge logarithmically as they
are approached.

\vskip1pc What happens when the point of inversion is changed is
also interesting. By translating this point along the  axis of
symmetry we break the up-down symmetry. The discocyte morphs
continuously into a stomatocyte. It is also possible to describe
this transition analytically. Surface areas, however, are not
preserved under inversion. To follow the transition in a physically
meaningful way, the geometries should therefore be rescaled so that
they possess the same area. One may then track the volume as a
function of the external force tethering the poles together. The
maximum external force consistent with this constraint occurs for
the symmetric discocyte; a higher value presumably would rupture
the membrane. For each value below this maximum, there is a unique
axially symmetric geometry. As the force is reduced, the discocyte
first inflates asymmetrically until a point is reached where it is
identifiable as a stomatocyte; thereafter it deflates into an
inverted sphere within a sphere. We have not conducted a complete
stability analysis; however, there are reasons to expect this
description to be stable with respect to small perturbations
breaking the axial symmetry.


\vskip1pc The paper is organized as follows:  in section 2 we
describe inverted minimal surfaces focusing on the stresses within
them; in section 3 we provide a detailed physical interpretation of
the inverted catenoid. Some of these results were announced in
\cite{guvencastro}.  We finish with a brief discussion and an
outline of our plans for future work.

\section{\Large \sf Willmore energy and inverted minimal surfaces}

A parametric description of an embedded two-dimensional surface in
three-dimensional space is provided by the mapping $(u^1,u^2) \to
{\bf X}(u^1,u^2)$. The Willmore energy associated with this surface
is given by
\begin{equation}
H\left[{\bf X}\right]= \int dA~ \left(K^{ab} -  \frac{1}{2}g^{ab} K
\right) \left(K_{ab}- \frac{1}{2} g_{ab} K\right)\,.
\label{Willmore}
\end{equation}
where  $g_{ab}$ is the metric, $K_{ab}$ is the extrinsic curvature
tensor, and $K=g^{ab} K_{ab}$ is its trace (twice the mean
curvature). $dA$ is the induced element of area. This notation is
summarized in Appendix I. $H[{\bf X}]$ is a measure of the energy
associated with bending. It has the remarkable property that it is
invariant with respect to the conformal transformations of the
ambient space \cite{Willmore}: thus $H[{\bf X}]$ is invariant not
only under the transformations of the surface induced by the
Euclidean motions, translations and rotations, it is also invariant
under transformations preserving angles: dilations ${\bf X}\to
\lambda {\bf X}$ and inversion,
\begin{equation} {\bf X}\to \frac{{\bf X}}{|{\bf X}|^2}\,.
\end{equation}

\vskip1pc Modulo a topological contribution proportional to the
Gauss-Bonnet invariant $H$ coincides with the Helfrich energy
\begin{equation}
H\left[{\bf X}\right]=\frac{1}{2}\int dA~K^{2}\,.
\end{equation}
There is only one local two-dimensional bending energy.

\vskip1pc The equilibrium surfaces of the energy (\ref{Willmore})
satisfy the Euler-Lagrange equation
\begin{equation}
  -\, \nabla^2 K + {1\over 2}
 \left( K g_{ab}- 2 K_{ab}\right) K^{ab}  K =0\,, \label{shapeequation}
\end{equation}
where $\nabla^2$ is the surface Laplacian compatible with the metric
$g_{ab}$.  It is clear that minimal surfaces, satisfying $K=0$ are
solutions of Eq.(\ref{shapeequation}). However, because of the
conformal symmetry  of the shape equation, these surfaces map under
inversion to new solutions of Eq.(\ref{shapeequation}). This is true
not just infinitesimally but also for finite conformal
transformations. In particular, it is true for inversion.

\vskip1pc Let us first examine the behavior of minimal surfaces
under inversion.  In particular, let us first identify the equation
satisfied by the transformed surface. It is well known that the
principal curvatures transform under inversion by $C_I \to \bar
C_I$, where (see, for example, \cite{Gray})
\begin{equation}
\bar C_I = |{\bf X}|^2 \left( C_I- 2 \,{({\bf X}\cdot {\bf n})\over
|{\bf X}|^2}\right)\,.
\end{equation}
Thus the transformed mean curvature $K= C_1+C_2$ is given by
\begin{equation}
\bar K= |{\bf X}|^2 \left( K - 4 \,{({\bf X}\cdot {\bf n})\over
|{\bf X}|^2}\right)\,. \label{eq:KKbar}
\end{equation}
We conclude from Eq.(\ref{eq:KKbar}) that a surface satisfying
\begin{equation}
 K= 4 \,{{\bf X}\cdot {\bf n}\over
|{\bf X}|^2} \label{eq:KXn}
\end{equation}
is mapped under inversion to a minimal surface, and conversely. It
is clear that the only minimal surfaces mapping to minimal surfaces
are the planes through the origin.\footnote{\sf Eq. (\ref{eq:KXn})
can also be written as ${\bf n}\cdot \left[\nabla^2 {\bf X} + 4 {\bf
X}/|{\bf X}|^2\right]=0$. Note, however, that there are no solutions
consistent with peeling off ${\bf n}$. Surfaces satisfying this
equation can also be interpreted  as the stationary configurations
of the functional $H_1[{\bf X}] = \int {dA/|{\bf X}|^{4}}$. } If the
origin itself lies on the surface, $K$ will diverge there unless the
surface aligns along ${\bf X}$ sufficiently fast as $|{\bf X}|\to
0$. While this is true if the original minimal surface has planar
ends it is not true if the ends behave like a catenoid. Inverted
minimal surfaces typically possess curvature singularities.

\vskip1pc Inverted minimal surfaces also satisfy Eq.
(\ref{shapeequation}).
Thus any solution to Eq.(\ref{eq:KXn}) is also a solution to
Eq.(\ref{shapeequation}). It is straightforward, but instructive, to
confirm this fact with an explicit calculation: we note that, modulo
Eq. (\ref{eq:KXn}), we have
\begin{equation}
 \partial_a K= {4\over |{\bf X}|^2 } (K_{ab}- {1\over
2} g_{ab} K) \, ({\bf X}\cdot {\bf e}^b)\,. \label{eq:parK}
\end{equation}
This makes use of the Weingarten equations, $\partial_a {\bf n}=
K_{ab}\, {\bf e}^b$. Eq.(\ref{shapeequation}) follows (except
possibly at the origin) on using the Gauss equations, $\nabla_a {\bf
e}_b = -K_{ab}\, {\bf n}$ as well as the contracted Codazzi-Mainardi
equations,
\begin{equation}
\nabla_a K^{ab} - \nabla^b K=0\,. \label{eq:CM}
\end{equation}
Here $\nabla_a$ is the surface covariant derivative compatible with
$g_{ab}$. Unlike minimal surfaces, the equilibrium of the inverted
counterpart involves a non-trivial cancelation between the Laplacian
of $K$ and the terms cubic in $K_{ab}$. If the origin lies on the
surface, however, equation (\ref{eq:KXn}) will not generally hold at
this point and there will be a geometrical singularity there. In the
following section we will show how to interpret this apparent
pathology physically in terms of a localized external force acting
on the membrane at this point.

\subsection{\Large \sf Conserved stress tensor and Noether charges
for the inverted minimal surfaces}

\vskip1pc The physical interpretation of geometrical singularities
in the transformed geometry at the origin of inversion is
facilitated by identifying a stress tensor with the geometry. This
tensor is given in terms of the extrinsic curvature tensor by
\cite{Stress,Auxiliary}
\begin{equation}
{\bf f}^a = K (K^{ab}- {1\over 2} g^{ab} K) \,{\bf e}_b -\partial^a
K\, {\bf n}\,. \label{eq:stressdef}
\end{equation}
In equilibrium, ${\bf f}^a$ is conserved so that $\nabla_a \,{\bf
f}^a =0$. It is simple to check that the local conservation law
reproduces the shape equation (\ref{shapeequation}).

\vskip1pc The transformed stress tensor is given by
\begin{equation}
\bar{\bf f}^a= |{\bf X}|^4 \left(|{\bf X}|^2 \, {\bf f}^a + 4
\left(K^{ab}- {1\over 2} g^{ab} K\right)\, {\bf f}_{b\,0}\right)\, ,
\label{eq:finvgen}
\end{equation}
where
\begin{equation}
{\bf f}_0^a = ({\bf e}^a\cdot {\bf X}) \, {\bf n} -({\bf n}\cdot
{\bf X}) \, {\bf e}^a\,. \label{eq:f0def}
\end{equation}
The role of ${\bf f}_0^a$ in the context of conformal symmetry has
been noted elsewhere \cite{ConfJ}. Curiously, it also provides an
effective surface stress tensor for any external Laplace force
\cite{Laplace}.

\vskip1pc In a minimal surface with $K=0$ the stress vanishes.
However, it does not vanish in its inverted counterpart satisfying
Eq. (\ref{eq:KXn}). The corresponding stress tensor is given by
\begin{equation} {\bf f}^a= - {4\over |{\bf X}|^2}
\left(K^{ab}- {1\over 2} g^{ab} K\right)\, {\bf f}_{b\,0}\,.
\label{eq:finv}
\end{equation}
It is also straightforward to demonstrate that ${\bf f}^a$ given by
Eq. (\ref{eq:finv}) is conserved if ${\bf X}\ne 0$: one uses Eq.
(\ref{eq:parK}) as well as the identity,
\begin{equation}
\nabla_b {\bf f}_{0\,a}= g_{ab}\,{\bf n} + K_{b}{}^c (({\bf X}\cdot
{\bf e}_a)\, {\bf e}_c - (a\leftrightarrow c))\,,
\end{equation}
satisfied by ${\bf f}^a_0$.

\vskip1pc As noted in \cite{guvencastro}, a curvature singularity at
a point will be associated with a source of stress at that point.
Let $\Gamma$ be any closed contour on the surface of the membrane
encircling the point. Stoke's theorem applied to the conservation
law implies that the closed line integral
\begin{equation}
\oint ds \, l^a {\bf f}_a \label{eq:lineint}
\end{equation}
is a constant vector ${\bf F}$ along contours that are homotopically
equivalent to $\Gamma$ on this surface \cite{MDG}. Here ${\bf l}=
l^a {\bf e}_a$ is the normal to $\Gamma$ tangent to the surface, and
$ds$ is the element of arc-length along $\Gamma$. The set of values
of ${\bf F}$ are the Noether charges associated with the translation
invariance of the energy.

\vskip1pc If the conservation law is valid everywhere, ${\bf F}$
must vanish on any topologically trivial loop. This is the case for
a membrane with spherical topology. Thus, if ${\bf F}\ne 0$ on such
a loop, there must be a source of stress within it \cite{MDG}. In
particular, a distributional source of stress manifest itself in a
curvature singularity which is picked up by the line integral. ${\bf
F}$ is a reparametrization invariant measure of the strength of
singularity.

\vskip1pc Note also that, in equilibrium, the total external force
acting on the membrane must vanish. In an axially symmetric membrane
the forces operating at the poles must be equal and opposite.

\vskip1pc A concrete implementation of these ideas will be developed
in the following section.

\section{\Large \sf Inversion of catenoids}

\vskip1pc Conformal invariance permits a remarkably simple
construction of a two-parameter family of axially symmetric
solutions of Eq. (\ref{shapeequation}).
Begin with a catenoid which solves the shape equation trivially as a
minimal surface with $K=0$.

\vskip1pc For a fixed axis of rotation, a catenoid
$\Sigma_{(R_0,\xi_0)}$, is given as the level set $\Phi(R,Z)=0$ of
the function \footnote{\sf Scaling the catenoid with an inverse
length will give an inverted geometry with the `correct'
dimensions.}
\begin{equation}
\Phi(R,Z)= R- {1\over R_0}\cosh ( R_0 Z + \xi_0)\,.
\end{equation}
It is characterized by a scale $1/R_0$ and an offset $\xi_0/R_0$
along the axis. Thus any two catenoids are related by scaling and
translation. They are the only non-trivial rotational minimal
surfaces.

\begin{figure}[h]
\begin{center}
\includegraphics[width=10cm]{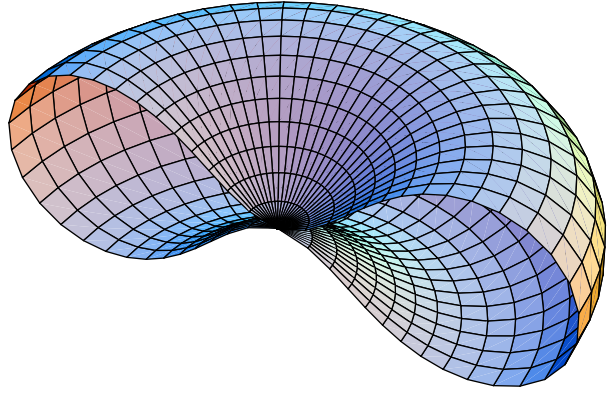}
\caption{{\small Section through symmetric biconcave discocyte}}
\end{center}
\label{fig1}
\end{figure}

\vskip1pc The image of the catenoid $\Sigma_{(R_0,\xi_0)}$ under
inversion in the origin is the surface $\bar \Sigma_{(R_0,\xi_0)}$
described as the level set
\begin{equation}
{R\over R^2 +Z^2} - {1\over R_0 }\, \cosh \left({R_0 Z\over (R^2
+Z^2)}+ \xi_0\right) =0\,. \label{eq:incat}
\end{equation}
This is a simple transcendental equation in the variables $R$ and
$Z$.

\vskip1pc One particular solution, the symmetric biconcave discocyte
with $\xi_0=0$, is illustrated in Fig.(1). The geometric profiles
for several different values of $\xi_0$ are displayed in Fig.(2).
Surface area is not preserved under inversion; thus, as we discuss
below, these profiles have been rescaled so that they all possess
the same surface area.
\begin{figure}[h]
\begin{center}
\includegraphics[width=5cm]{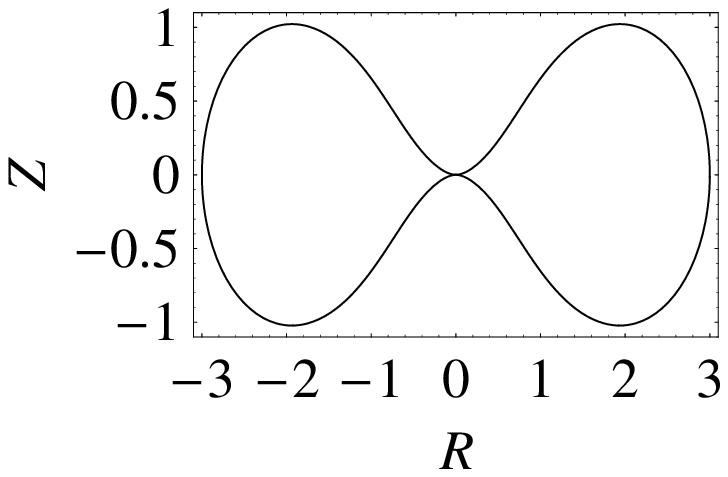}~~~~
\includegraphics[width=5cm]{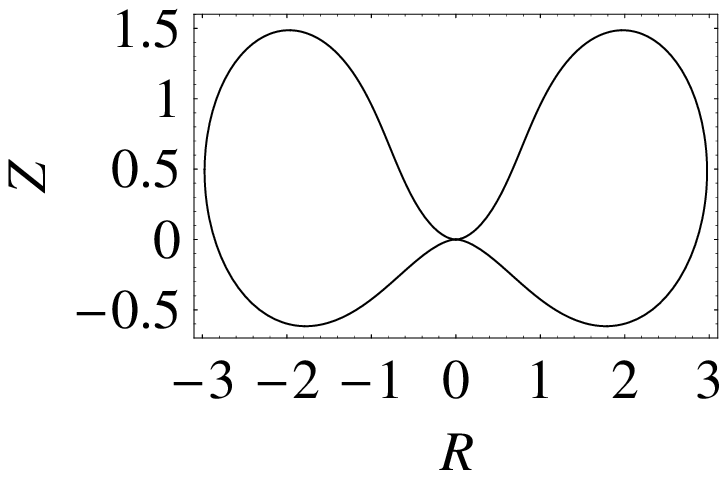}~~~~
\includegraphics[width=5.2cm]{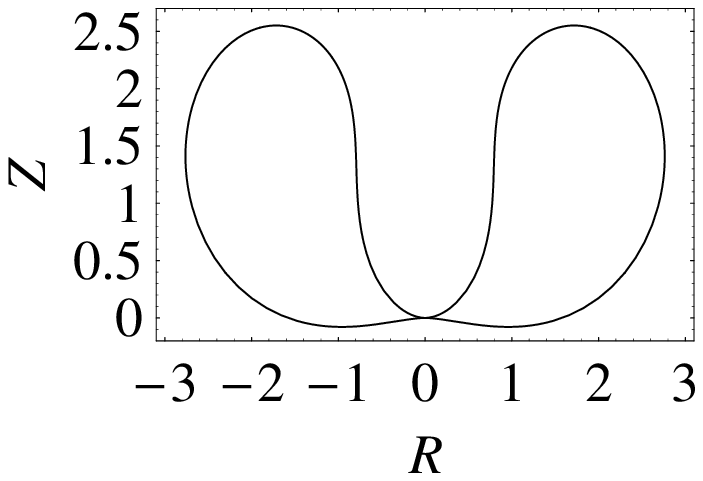}\\
\includegraphics[width=5.2cm]{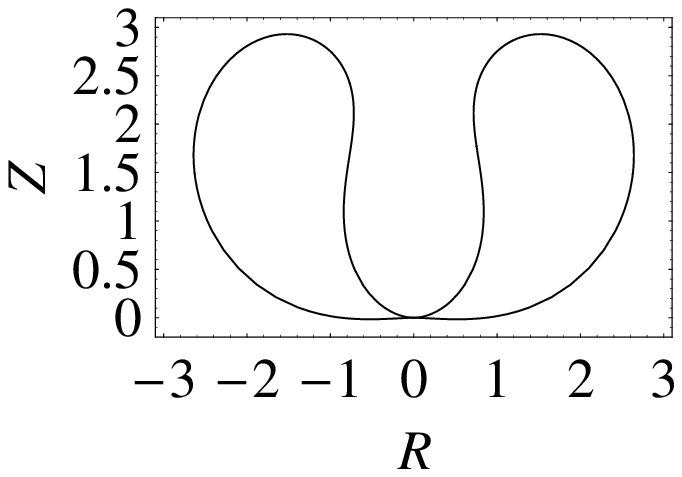}~~~~
\includegraphics[width=5.2cm]{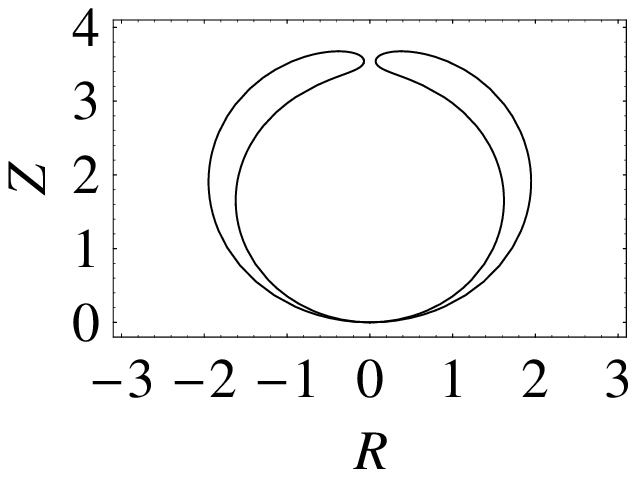}~
\includegraphics[width=5.4cm]{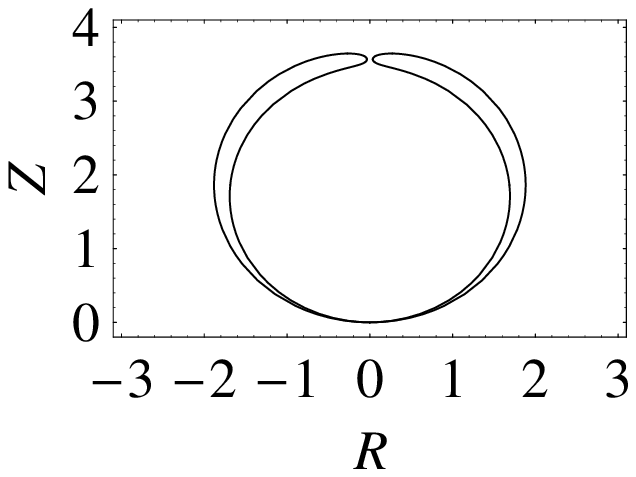}
\caption{{\small Geometric profiles for $\xi_{0}=0,1,2,3,50,100$.
The surface is generated by rotating the profile about the vertical
axis. The area is normalized.}}
\end{center}
\end{figure}
\\
\\
We note the following properties of these new surfaces:

\begin{itemize}
\item{} The geometry is bounded for all values of
$\xi_0$. The circle $R=1/R_0 \cosh \xi_0$ on the plane $Z=0$ is
mapped to the circle $R=R_0/\cosh \xi_0$ on the same  plane. The
remote regions of the catenoid map to the point of inversion. Thus
the north and south poles of this geometry touch. As we will see,
they do so with a common tangent plane $Z=0$. In this sense, the
inverted surface does not self-intersect. The topology is spherical.
However, there will be a singularity at the origin.

Generically, the behavior at the origin will not be so simple.
Consider, for example, the fate of another simple geometry, the
hyperboloid of revolution, under inversion in the origin.
Asymptotically, the hyperboloid is conical. Because cones map to
cones under inversion in the apex, the inverted geometry will be
conical at the origin; it is not spherical. The existence of a
tangent plane in the inverted catenoids is a consequence of the
faster asymptotic growth rate (exponential) of the catenoid.

\item{} When $\xi_0=0$, the surface is a biconcave
discocyte possessing up-down symmetry. This symmetry is a
consequence of the symmetry of the original catenoid with respect to
the point of inversion.

\item{} The symmetry is broken by translating the
point of inversion along the axis. Beyond some critical value,
$\xi^{*}_{0}=1.86823$ discussed below, the lower concave region
becomes vanishingly small: there is a transition from the symmetric
biconcave discocyte to an asymmetrical stomatocyte.

\item{} In the limit $\xi_0\to \infty$, the stomatocyte degenerates into
a completely deflated geometry consisting of a sphere within a
sphere connected by an infinitesimal neck, a catenoid. It is
possible to describe the complete sequence from discocyte to
limiting deflated stomatocyte analytically.
\end{itemize}

\noindent  It should be remarked that, technically, the lower
concave region persists for all values of $\xi_{0}$; the ratio
$\xi_{0}/R_{0}$ determines the position of the center along the
axis; thus no matter how large a value we take for $\xi_{0}$, there
always will be a region of the catenoid where $Z$ is negative; this
region will map to negative $Z$ in the inverted geometry and will
contain a concave component. Of course, as the profiles above
clearly illustrate, while  this concave region is always present, it
becomes small in comparison with the overall size of the geometry.
How small will be described quantitatively in our discussion of the
isoperimetric ratio.

\subsection{\Large \sf The isoperimetric ratio of the inverted shapes}

\vskip1pc To examine the geometry quantitatively, it is useful to
introduce a parametric representation of the catenoid in terms of
the angle that the tangent to the meridian makes with a plane of
constant $Z$:
\begin{equation}
 R\left(\Theta\right)= {1\over R_0} \csc\Theta\,,\quad Z
 \left(\Theta\right)= {1\over R_0}\ln \tan\Theta/2 +
{\xi_0\over R_0}\,.
\end{equation}
where $\Theta$ lies in the interval $\left[0,\pi\right]$. This angle
also provides a parametrization of the inverted catenoid
$\bar\Sigma$:
\begin{equation}
\bar R(\Theta)= {R_0\sin\Theta \over 1 + \sin^2 \Theta
(\ln\tan\frac{\Theta}{2} + \xi_0)^2}\,, \quad \bar Z(\Theta) =
{R_0\sin^2\Theta (\ln  \tan\frac{\Theta}{2} + \xi_0)\over 1 + \sin^2
\Theta (\ln \tan\frac{\Theta}{2} + \xi_0)^2} \,. \label{eq:barRZ}
\end{equation}
The interval of $\Theta$ does not change under inversion. A word of
caution: $\Theta$ and its inverted counterpart $\bar \Theta$ are not
the same angle i.e. $\Theta$ has nothing to do with the angles which
are preserved under conformal transformation \footnote {\sf More
generally, $\bar {\bf e}_a\cdot{\bf k}\ne {\bf e}_a\cdot {\bf k}$,
where ${\bf k}$ is a fixed unit vector.}.  While it might be more
natural to parametrize the inverted surface by $\bar\Theta$, the
advantage has to be weighed against the complicated form of the
corresponding functional forms of $\bar R$ and $\bar Z$.

\vskip1pc A straightforward calculation shows that the
parametrization by $\Theta$ is isothermal. The metric on the
inverted catenoid is described by the line element
\begin{equation}
ds^2=R_0^2
\,\Omega^2\,\left(\Theta\right)\left[d\Theta^2+\sin^2\Theta\,
d\varphi^2\right]\,,
\end{equation}
where $\varphi$ is the polar azimuthal angle, and the conformal
factor $\Omega$ is given by
\begin{equation}
\Omega\left(\Theta\right)^{-1}=1+
\sin^2\Theta\left(\log\tan\frac{\Theta}{2}+\xi_{0}\right)^2\,.
\end{equation}

\vskip1pc\noindent The area of the surface is given with respect to
this parametrization by
$A=4\pi R_0^2 {\cal A}(\xi_0)$
where
\begin{equation}
{\cal A}\left(\xi_0\right)=\frac{1}{2}\,\int^{\pi}_{0}d\Theta\,
\sin\Theta\, \Omega\left(\Theta\right)^2\,,
\end{equation}
with  range $\left(0,\infty\right)$.
The area of the surface
is used to express $R_{0}$ as a function of $\xi_{0}$ for a fixed
area $A$ as $R_{0}\left(\xi_0\right)=\sqrt{A/4\pi {\cal
A}\left(\xi_0\right)}$
The volume enclosed by the surface is given by
$V=\frac{4\pi R_0^3}{3}\, {\cal V} (\xi_0) \,,$
where
\begin{equation}
{\mathcal V}(\xi_0)= {1\over 2} \int^{\pi}_{0} d\Theta\,
\sin^3\Theta\left(1+\cos\Theta
\left(\log\tan\frac{\Theta}{2}+\xi_{0}\right)\right)\,\Omega^3(\Theta)\, ,
\end{equation}
and thus the isoperimetric ratio $\nu$, defined by
$\nu=V/V_{A}=3\sqrt{4\pi}V/A^{\frac{3}{2}}$ where $V_A$ is the
volume of a sphere of surface area $A$, is given by
\begin{equation}
\nu\left(\xi_{0}\right)={\cal V} \left(\xi_{0}\right)/{\cal
A}\left(\xi_{0}\right)^{\frac{3}{2}}\,.
\end{equation}
The maximum value, $v=1$ is attained by a sphere. Clearly the
constraint that the poles touch introduces bending energy. As a
consequence the value $v=1$ is not attained in any equilibrium
geometry with touching poles.

\vskip1pc The  isoperimetric ratio is plotted as a function of
$\xi_0$ in figure 3(a). $v\left(\xi_{0}\right)$ possesses three
extrema, two maxima and one minimum. The maxima $v_{\rm Max} =
0.66564$ occur when  $ \xi^*_{0}=\pm 1.86823$. The corresponding
geometry marks the transition from discocyte to stomatocyte
(depicted in figure 2). The shallow local minimum $v_{\rm
Min}=0.647639 $ occurs when $\xi_0=0$, the symmetric biconcave
geometry. We observe, in particular, the interesting fact that the
maximally inflated geometry is not the symmetric discocyte. $\nu$
vanishes asymptotically in the limits $\xi_0\to\pm\infty$. This
property is evident in the corresponding profile (see figure 2).

\vskip1pc We observe, in particular, that $v(\xi_0)$ may not be
inverted for $\xi_0$ on the interval $\xi_0 \in [0,\infty)$. This
indicates that it cannot be used in place of $\xi_0$ as the `order
parameter' to describe the discocyte-stomatocyte transition. As we
will see, the appropriate physical parameter is the Noether charge
or external force ${\bf F}$. This physical interpretation will be
developed in the following sections.

\vskip1pc\noindent
\begin{figure}[h]
\begin{center}
\includegraphics[width=7cm]{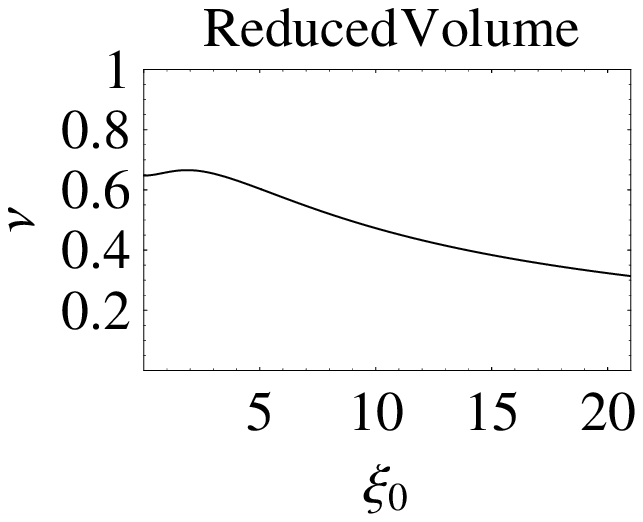}~~~~~
\includegraphics[width=7cm]{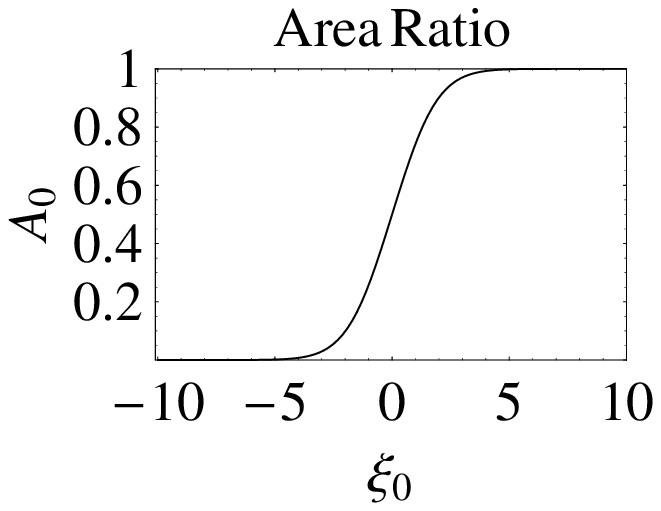}
\caption{{\small (a) Isoperimetric ratio vs $\xi_0$; (b) Area ratio
vs $\xi_0$. }}
\end{center}
\end{figure}\\

\vskip1pc It was pointed out in the previous section that, from a
technical point of view, the geometry remains biconcave for all
values of $\xi_{0}$. We will now demonstrate that one of the concave
regions becomes vanishingly small when $\xi_{0}$ is larger than the
critical value $\xi^{*}_{0}=1.86832$. To do this, we determine the
fraction of surface area lying above $Z=0$. We thus define the ratio

\begin{equation}
\frac{A_{0}}{A}=\frac{1}{\mathcal{A}\left(\xi_{0}\right)}\int^{\pi}_{\Theta_{0}
\left(\xi_{0}\right)}d\Theta\, \sin\Theta\, \Omega
\left(\Theta\right)\,,
\end{equation}
where $\Theta_{0}\left(\xi_{0}\right)=2\arctan
\left(\exp\left(-\xi_{0}\right)\right)$ is the tangent angle along
the meridian where the surface intersects $Z=0$.  For example, when
$\xi_{0}=0$, $\theta_{0}\left(0\right)=\frac{\pi}{2}$ and the area
ratio is one half.  Figure 3 (b) illustrates clearly the rapid
approach of the ratio to $1$ above the critical value, $\xi_{0}^*$.

\subsection{\Large \sf Curvature singularities}

\vskip1pc We expect a curvature singularity at the poles. This is
because the only regular equilibrium geometries with spherical
topology are the round spheres. We will now confirm that the
geometries described by Eq. (\ref{eq:incat}) are regular everywhere
except at the poles where they display a (logarithmic) curvature
singularity. The strength of this singularity will be related to the
external force tethering the poles together.

\vskip1pc In the case of the symmetric discocyte it is very simple
to study this singular behavior analytically. To do this, note that,
in the neighborhood of the origin, Eq. (\ref{eq:incat}) is
approximated by
\begin{equation}
{1\over R} \approx {1\over 2R_{0}} \exp\left(R_{0}Z\over R^2\right)
\end{equation}
which can be inverted for $Z$ as a function of $R$,
\begin{equation}
 \frac{Z}{R_{0}} \approx -\left(\frac{R}{R_{0}}\right)^2\log\left(\frac{R}{2R_{0}}\right)
\end{equation}
when $R\approx 0$. Whereas $Z_{,R}$ vanishes at $R=0$, it is clear
that both the curvature along the meridian $C_\perp\approx Z_{,RR}$,
as well as that along the parallel, $C_\|\approx Z_{,R}/R$, diverge
logarithmically (see appendix II). Indeed, $C_\|$ and $C_\perp$
exhibit identical logarithmic divergences:
\begin{eqnarray}
C_\|,~~C_\perp&\approx& -\frac{2}{R_{0}}\log\left(\frac{R}{2R_{0}}\right)\nonumber
\label{eq:C0}
\end{eqnarray}
The poles are umbilical points of the geometry, albeit in a singular
way.

\vskip1pc We will now show that this qualitative behavior holds for
all values of $\xi_0$. However, a different strategy becomes more
appropriate when $\xi_0\ne 0$. We have the following exact
expressions for the principal curvatures with respect to the
parametrization given by (\ref{eq:barRZ})
\begin{eqnarray}
C_\perp &=&\frac{
\partial^2_{\Theta}R~\partial_{\Theta}Z-\partial^2_{\Theta}Z~\partial_{\Theta}R}
{\left(\left(\partial_{\Theta}R\right)^2
+\left(\partial_{\Theta}Z\right)^2\right)^{\frac{3}{2}}}\nonumber\\
&=& -\frac{1}{R_{0}}\left(3+2\cos\Theta
\left(\log\tan\frac{\Theta}{2}+\xi_{0}\right)+
\sin^2\Theta\left(\log\tan\frac{\Theta}{2}+\xi_{0}\right)^2\right)\,,
\end{eqnarray}
and
\begin{eqnarray}
C_\|&=&-\frac{\partial_{\Theta}Z}{R\left(\left(\partial_{\Theta}R\right)^2
+\left(\partial_{\Theta}Z\right)^2\right)^{\frac{1}{2}}}\nonumber\\
&=&-\frac{1}{R_{0}}\left(1+2\cos\Theta\left(\log\tan\frac{\Theta}{2}
+\xi_{0}\right)-\sin^2\Theta\left(\log\tan\frac{\Theta}{2}
+\xi_{0}\right)^2\right)\,.
\end{eqnarray}

\vskip1pc\noindent It is clear, by inspection, that the two
principal curvatures diverge logarithmically at the poles $\Theta=0$
and $\Theta=\pi$. Note, however, that  their difference remains
finite with the value $C_\perp- C_\| =-2/R_{0}$ as consistency with
Eq.(2) of \cite{guvencastro} demands. The angular dependence of the
curvatures for different values of $\xi_0$ is plotted in Figures 4
and 5.

\begin{figure}[h]
\begin{center}
\includegraphics[width=5.2cm]{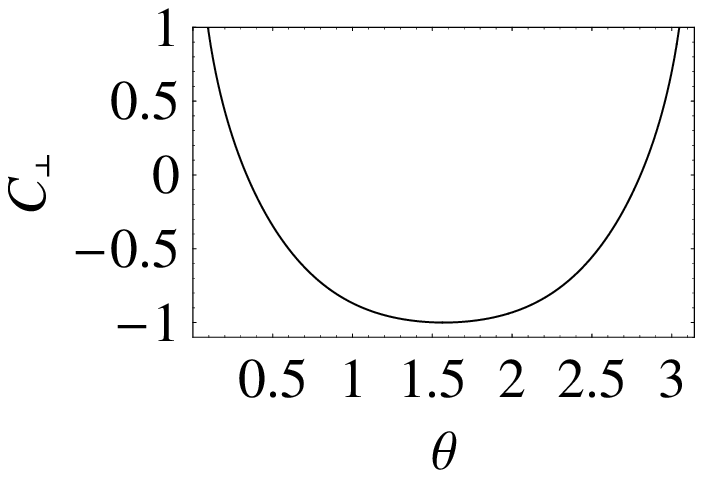}~~~~
\includegraphics[width=5.2cm]{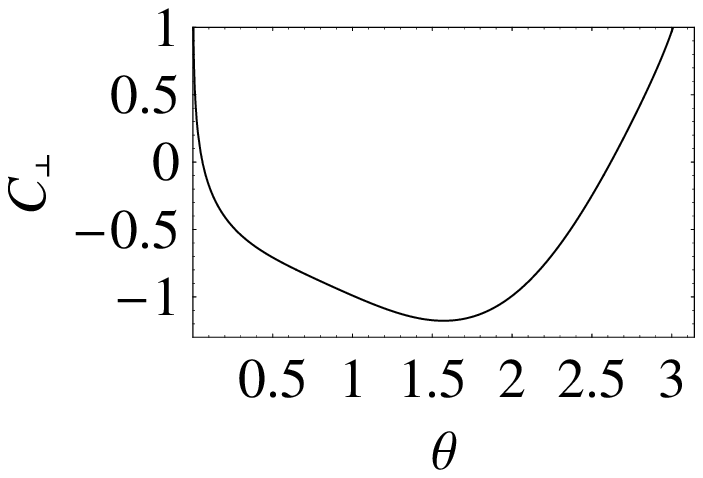}~~~~
\includegraphics[width=5.2cm]{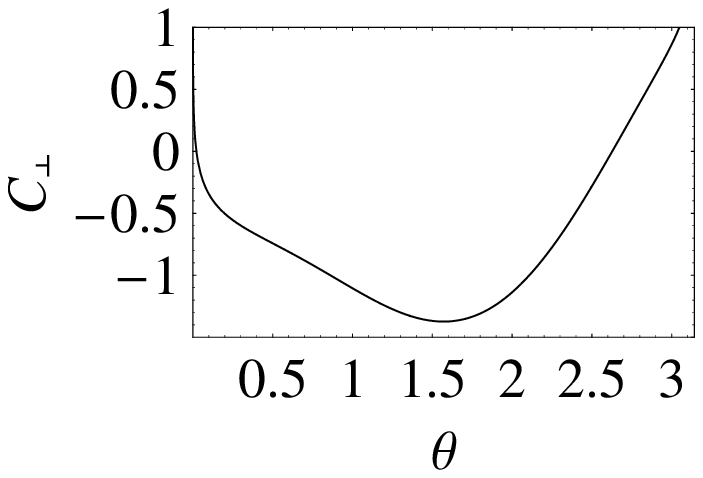}
\caption{{\small $C_\perp$ vs. $\Theta$ for three different values
of $\xi_0$: $R_{0}=3$, $\xi_{0}=0$ (left);
$R_{0}=3\sqrt{\mathcal{A}\left(0\right)/\mathcal{A}\left(2\right)}$, $\xi_{0}=2$
(center); $R_{0}=3\sqrt{\mathcal{A}\left(0\right)/\mathcal{A}\left(3\right)}$,
$\xi_{0}=3$ (right). }}
\end{center}
\end{figure}

\begin{figure}[h]
\begin{center}
\includegraphics[width=5cm]{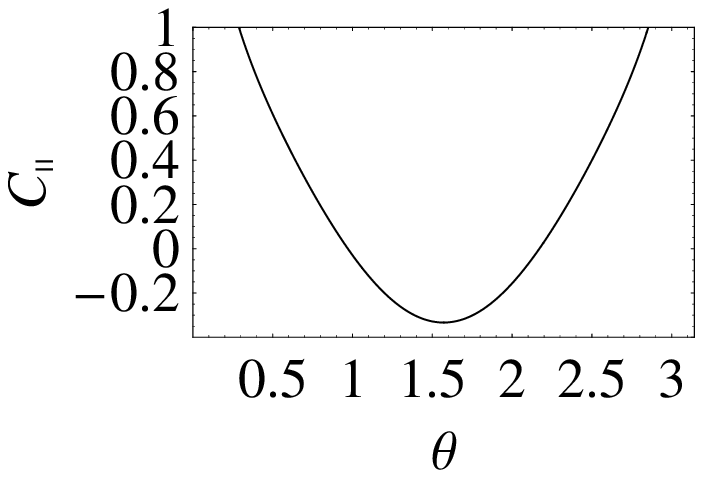}~~~~
\includegraphics[width=5cm]{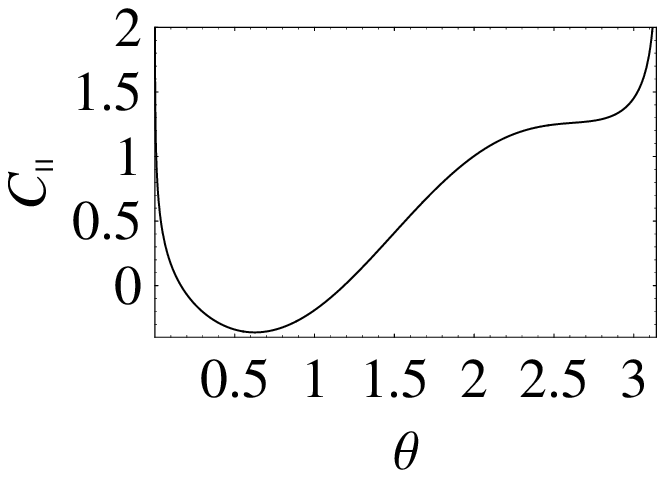}~~~~
\includegraphics[width=5cm]{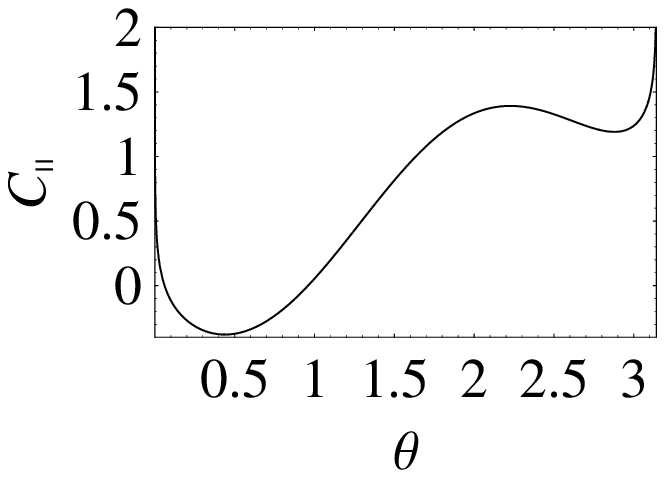}
\caption{{\small $C_\|$ vs. $\Theta$: $R_{0}=3$, $\xi_{0}=0$ (left);
$R_{0}=3\sqrt{\mathcal{A}\left(0\right)/\mathcal{A}\left(2\right)}$, $\xi_{0}=2$
(center); $R_{0}=3\sqrt{\mathcal{A}\left(0\right)/\mathcal{A}\left(3\right)}$,
$\xi_{0}=3$ (right). }}
\end{center}
\end{figure}

\vskip1pc\noindent Note, in particular, how sharp the divergence is
in the neighborhood of the south pole. Note also that both
curvatures diverge to infinity. This peculiarity is
associated with the logarithm; it is not apparent visually in the
corresponding profiles due to the existence of a tangent plane
(figure 2).

\vskip1pc The singularity at the poles indicates that these
surfaces, unlike the catenoid, require external forces along the
axis of symmetry to support them.

\vskip1pc It should be pointed out that the characteristic
logarithmic singularity we have examined is, in fact,  predicted by
the linear theory. Consider the linear approximation to the Monge
representation of the surface in terms of its height $h$ above a
plane. If the gradients of $h$ are small, the bending energy $H$ can
be expressed as
\begin{equation}
 H = {1\over 2} \int d^2 x \, (\nabla_\perp^2 h)^2  + {\cal O}(4)
\end{equation}
where $\nabla_\perp$ is the gradient on the plane.  The
corresponding Euler-Lagrange equation is the biharmonic equation
equation on the plane, $-(\nabla_\perp^2)^2 \, h=0$. It is
straightforward to show that the most general axially symmetric
solution of this equation is given by
\begin{equation}
h = c_0 + c_1 R^2 \log R^2 + c_2 \log R + c_3 R^2
\end{equation}
The linear Monge approximation to the inverted catenoids is given by
solutions with $c_0,c_2=0$. Of course, it is beyond the scope of
this approximation to model their global behavior.

\subsection{\Large \sf Singularities as manifestations of sources of stress}

\vskip1pc Axial symmetry dictates that the force on the membrane
must be directed parallel to the axis, ${\bf F}= -2\pi c{\bf k}$.
We have shown recently that $c$ is the constant appearing in Eq. (2)
of \cite{guvencastro}. Thus using this equation and Eq.
(\ref{eq:KXn}) we find that an inversion of the catenoid must
satisfy the following equation
\begin{equation}
C_{\perp}-C_{\|}=\frac{c|X|^2}{2R^2}\,.
\end{equation}
If we now use the explicit expressions for the principal curvatures
we identify $ c=-4/R_{0}$. This is a reparametrization invariant
measure of the singularity. It is also possible to see this directly
using the expression given in Eq. (\ref{eq:lineint}): without loss
of generality, choose the contour to be a closed circle of polar
radius $R$ encircling the north pole. Now ${\bf l}$ is the tangent
to the meridian and $ds = R d\phi$. We do not use Eq.(\ref{eq:finv})
which is ill-defined at the pole. Using (\ref{eq:stressdef}) and
(\ref{eq:C0}), and that on the north pole ${\bf n}= -{\bf k}$, we
obtain
\\
\begin{eqnarray}
l_a {\bf f}^a &=& {1\over 2} (C_\perp^2 - C_\parallel^2) \, {\bf l}
- (C_\perp + C_\parallel)' \, {\bf n}\nonumber\\
&\approx& (C_\perp + C_\parallel)' \, {\bf k}\nonumber\\
&=& {4 R_0\over R}\,\, {\bf k}\,.
\end{eqnarray}
As a consequence,
\begin{equation}
{\bf F}= {8\pi R_0}\,\,{\bf k}\,,
\end{equation}
so that $c=-4/R_0$. The sign indicates that the force is always
directed towards the interior.

\begin{figure}[h]
\begin{center}
\includegraphics[width=12cm]
{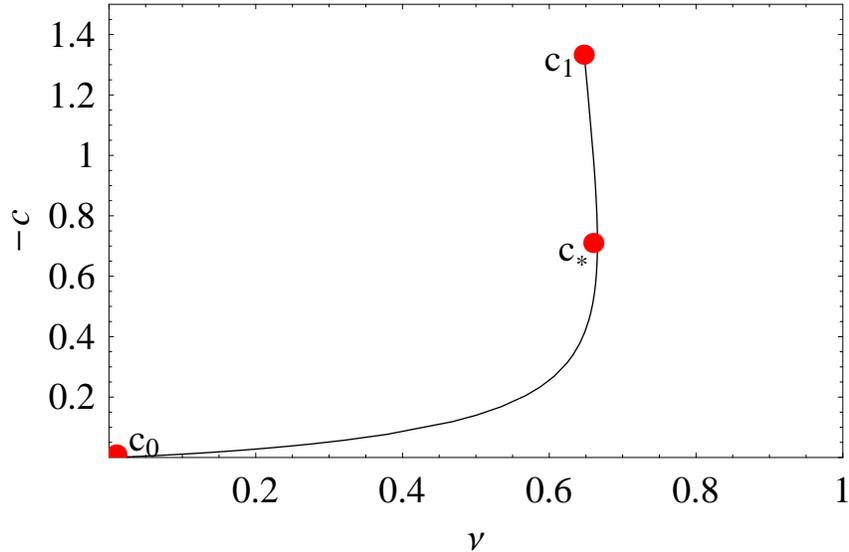}
\caption{{\small External force $c$ vs. v}}
\end{center}
\end{figure}

\subsection{\Large \sf The Phases of a tethered membrane}

\vskip1pc We now possess a physical interpretation of the inverted
geometry as an equilibrium with a pair of equal and opposite
localized external forces applied at the poles holding them
together. This might be compression applied outside, or the tension
supplied by an internal tether. The curvature singularity at the
poles is a manifestation of local forces. It is unrelated to the
fact that they touch: they touch simply because all distant parts of
the original catenoid get mapped to the origin, a peculiarity of the
exact solution we have been handed. In a subsequent publication we
will construct equilibrium configurations with a fixed non-vanishing
interpolar distance.

The axial force $c$ may be interpreted as the order parameter
describing the  transition from discocytes to stomatocytes.
The axial force and the reduced volume may be used to characterize
axially symmetric compact shapes. The geometries with tethered poles
that we have considered define a finite trajectory on this space
illustrated in figure 6.  Each point on this trajectory represents a
specific geometry: the discocyte lies at one end (with the maximum
fore $c_{1}=-4/3$), the limiting geometry consisting of a sphere
within a sphere at the other (vanishing force $c_{0}=0$).
The maximally inflated equilibrium geometry ($c_{*}=-0.711923$)
provides a natural division of  the trajectory into two phases:
discocytes and stomatocytes.

\section{\sf Discussion}

The invariance of the bending energy of two dimensional surfaces
with respect to inversion in any point has been used to obtain exact
results for a family of deflated biconcave equilibrium geometries
with spherical topology and with tethered poles. The external force
tethering the poles is imprinted as a curvature singularity on the
membrane geometry.  We end up with a geometry describing a
completely different physical setup from the catenoid geometry we
began with.

\vskip1pc There is a, of course, a superficial similarly to the
discocyte-stomatocyte transition of fluid membranes induced by a
change in the bilayer asymmetry.
The quantitative study of the discocyte shape of a deflated fluid
membrane and its stability has a long history dating back to Deuling
and Helfrich's pioneering work in the seventies \cite{DH} followed,
after a long pause, by a series of ground-breaking studies by
several groups (see, for example,
\cite{SvetinaSeifertBerndlLipowsky} as well as
\cite{Seifert,Lipowsky,Svetina} for reviews). The transition
explored in this work is associated with a change in the bilayer
asymmetry. An energy penalty or constraint on the area difference is
introduced. These constraints break the conformal symmetry of the
bending energy pointwise.  In the inverted catenoid, on the other
hand, there is no bilayer asymmetry. The transition is associated
with a change in the localized external forces. The action of these
forces breaks conformal symmetry only at isolated points. As a
consequence, it is possible to constrain the isoperimetric ratio
without sacrificing the conformal invariance of the energy.  While
it is tempting to emphasize the analogy between the two systems, it
is probably a mistake to dwell on it.

\vskip1pc These differences notwithstanding, the simple model we
have described is relevant to the study of biological membranes in
other ways. For the action of forces external to the membrane
controlling its shape is the rule rather than the exception in
biology. The cytoskeleton pushes and pulls on the plasma membrane
\cite{wortis}. Indeed, our toy model might after all be relevant to
the stability of the biologically relevant biconcave discocyte shape
assumed by a healthy erythrocite which involves this interaction in
an essential way.

\vskip1pc It is also possible that our model is relevant for the
theoretical study of the geometry of two very important organelles:
the Golgi complex and the mitochondrion. It is known that the Golgi
complex is not an isolated equilibrium structure \cite{Alberts}. It
is stabilized against breakup by a complex network of microtubules
that is dismantled and reassembled during mitosis. And the
convoluted inner membrane of the mitochondrion is likely to be under
compressive stress due to its confinement within the outer membrane
\cite{SanDiego}.

\vskip1pc In this context it is perhaps worth stressing that, in
principle, it is always possible to reconstruct the distribution of
external forces $\{{\bf F}_1,\dots,{\bf F}_N\}$ from a knowledge of
the geometry in the neighborhood of an appropriate set of contours.
The external forces leave their imprint on the local membrane
geometry. As imaging techniques improve, this could become a very
useful diagnostic tool in the study of intracellular membranes.

\vskip1pc The solutions we have described are very special in that
the poles touch. Intuitively, one would expect that solutions exist
with poles held a finite distance apart.  In the early nineties
Ou-Yang Yhong-can, Naito and Okuda identified an exact analogue  of
such configurations for a model with a finite spontaneous curvature
\cite{OuY}. Generally, however, the effect of an external force will
be to pull out two cylindrical tethers. Tether pulling in the
context of a physically more realistic Hamiltonian has been the
subject of some very nice recent theoretical analysis (see, for
example, \cite{Heinrich,Prost,Powers}). Such geometries will occur
as solutions of the axially symmetric shape equation we derived in
reference \cite{guvencastro}. A detailed description of the
solutions of this equation is, however, beyond the scope of this
paper and will be addressed in a forthcoming paper. It will come as
no surprise that conformal symmetry has a role to play in their
characterization.

\vskip1pc We have said nothing about the stability of the shapes we
have discussed.  Expanding the bending energy out to second order in
deformations about a  minimal surface, one can show that \cite{defo}
\begin{equation}
 H = {1\over 2} \, \int dA \,\Phi (-\nabla^2 + {\cal R})^2 \Phi\,,
 \label{eq:H2}
\end{equation}
where $\Phi= \delta {\bf X}\cdot {\bf n}$ is the normal deformation
and ${\cal R}$ is the scalar curvature (twice the Gaussian curvature
$C_1 C_2$). Remarkably, this expression depends only on intrinsic
geometry.  It should be compared with the corresponding expression
for the area,
\begin{equation}
 H = \int dA \,\Phi (-\nabla^2 + {\cal R}) \Phi\,.
\end{equation}
In general, for any minimal surface ${\cal R} < 0$. Thus the
catenoid spanning two rings becomes unstable as a minimal surface
beyond a certain maximum separation. By contrast, note that the
expression given by (\ref{eq:H2}) is manifestly positive. Thus, if
tension may be ignored, the same catenoid is stable as a Willmore
surface. The conformal invariance of the energy should hold order by
order in perturbations theory. This suggests that the inverted
shapes are also stable. While one does need to confirm that there is
no subtlety associated with singularities, conformal invariance
clearly simplifies the analysis of stability.

\vskip1pc So much for axially symmetric configurations. It is
clearly possible to generate non-axially symmetric geometries held
by pair of tethers by inverting these geometries in any point off
the axis of symmetry. It appears reasonable to conjecture that all
equilibrium shapes with a pair of tethers are generated by
inversion.

\vskip1pc It is also clear, however, that there is much more to this
story.  For the catenoid is but the simplest of a vast and growing
number of interesting minimal surfaces. How do we describe
equilibria held in place by three or more tethers? It is not
unreasonable to guess that the inversion of non-axially symmetric
minimal surfaces will play a role. We are currently examining the
properties under inversion of the natural generalization of the
catenoid, the n-noids of Jorge and Meeks \cite{JorgeMeeks}. This
could provide a non-perturbative handle on yet another problem of
interest in membrane biophysics: how does one identify the stable
configurations assumed by proteins when the interaction between them
is mediated by membrane curvature \cite{Kim}?


\vskip2pc
\appendix\noindent
 {\Large\bf Appendix I} \vskip1pc\noindent The tangent
vectors adapted to the parametrization are ${\bf e}_a =\partial_a
{\bf X}$ and ${\bf n}$ is the unit normal. The induced metric on the
surface and the extrinsic curvature are given by $g_{ab} ={\bf
e}_a\cdot{\bf e}_b$ and $K_{ab}= {\bf e}_a\cdot
\partial_b\, {\bf n}$ respectively \cite{Carmo,Ros,Spivak}.
Indices are raised with the inverse metric $g^{ab}$. The element of
area is given by $dA = \sqrt{{\rm det}\, g_{ab}}\, d^2 u$. $K$
denotes (twice) the mean curvature: $K= g^{ab}K_{ab}$

\vskip2pc
\appendix
\noindent {\Large\bf Appendix II} \vskip1pc\noindent
 The principal curvatures
are given by
\begin{eqnarray}
C_\perp = \bar\Theta'\,,\quad C_\| = {\sin \bar\Theta/\bar R} \nonumber
\end{eqnarray}
where $\bar\Theta$ denotes the angle that the tangent along the
meridian makes with the planes of constant $Z$, and the prime
denotes a derivative with respect to arc-length along the meridian.
The arc-length is given by
\begin{eqnarray}
ds^2  = (1 + Z_{,R}^2) dR^2\nonumber
\end{eqnarray}
Thus $\cos\bar\Theta = R'= -1/(1 + Z_{,R}^2)^{1/2}\approx -1$,
$\sin\bar\Theta = -Z'= Z_{,R}/(1 + Z_{,R}^2)^{1/2}\approx Z_{,R}$,
and $\bar\Theta' \approx Z_{,RR}$.

 \vskip3pc \noindent{\large \sf Acknowledgments}

\vspace{.5cm}

\vskip1pc\noindent  We have benefitted from conversations with
Markus Deserno, Martin M\"uller and Francisco Solis. Partial support
from CONACyT grant 51111 as well as DGAPA PAPIIT grant IN119206-3 is
acknowledged.

\end{document}